\documentclass[reprint,showpacs,amsmath,amssymb,aps,pra,superscriptaddress,nobibnotes]{revtex4-2}

\usepackage{graphicx}
\usepackage{dcolumn}
\usepackage{bm}
\usepackage{bbm}
\usepackage{amsthm}
\usepackage{mathtools}
\usepackage{physics}
\usepackage{binarytree}
\usepackage{tikz}
\usepackage{verbatim}
\usepackage{pgfplots}
\usepackage{placeins}
\usetikzlibrary{decorations.pathreplacing}
\usetikzlibrary{arrows.meta}
\usetikzlibrary{graphs}

\usepackage[colorlinks, linkcolor=red, anchorcolor=blue, citecolor=green]{hyperref}

\pgfplotsset{compat=1.16}
\begin{document}

\title[Deterministic quantum one-time pad via Fibonacci anyons]
  {Deterministic quantum one-time pad via Fibonacci anyons}

\author{Cheng-Qian Xu}

\affiliation{Institute of Physics, Beijing National Laboratory for
  Condensed Matter Physics,\\Chinese Academy of Sciences, Beijing
  100190, China}

\affiliation{School of Physical Sciences, University of Chinese
  Academy of Sciences, Beijing 100049, China}

\author{D. L. Zhou} \email[]{zhoudl72@iphy.ac.cn}

\affiliation{Institute of Physics, Beijing National Laboratory for
  Condensed Matter Physics,\\Chinese Academy of Sciences, Beijing
  100190, China}

\affiliation{School of Physical Sciences, University of Chinese
  Academy of Sciences, Beijing 100049, China}

\affiliation{Collaborative Innovation Center of Quantum Matter,
  Beijing 100190, China}

\affiliation{Songshan Lake Materials Laboratory, Dongguan, Guandong
  523808, China}

\date{\today}

\begin{abstract}
  Anyonic states, which are topologically robust because of their peculiar
  structure of Hilbert space, have important applications in quantum computing
  and quantum communication. Here we investigate the capacity of the
  deterministic quantum one-time pad (DQOTP) that uses Fibonacci anyons
  as an information carrier. We find that the Fibonacci particle-antiparticle
  pair produced from vacuum can be used to asymptotically send $2\log_2 d_{\tau}$
  bits of classical information ($d_{\tau}$ is the quantum dimension of a Fibonacci
  anyon $\tau$), which equals anyonic mutual information of the pair.
  Furthermore, by studying the DQOTP via a parameterized state of six Fibonacci
  anyons with trivial total charge, we give the analytical results of the
  maximum number of messages that can be sent for different parameters, which is
  a step function with every step corresponding to a regular simplex from the
  viewpoint of geometry. The results for the maximum number of messages sent by
  the DQOTP can be explained by anyonic accessible information.
\end{abstract}

% \keywords{Suggested keywords}
                              
\maketitle

\section{\label{sec:intro}Introduction}

In quantum communication the superdense coding~\cite{PhysRevLett.69.2881, PhysRevLett.76.4656}
shows that channel capacity can be increased by the entanglement between Alice
(sender) and Bob (receiver). Afterwards Schumacher and
Westmoreland~\cite{PhysRevA.74.042305} presented a quantum protocol called
the quantum one-time pad, which can be seen as a generalization of the superdense
coding. 
In the protocol, Alice and Bob share a quantum state $\rho^{AB}$. The
classical message, which Alice wants to send to Bob, is encoded by Alice's quantum operations
on her subsystem. Then Alice sends her subsystem to Bob, and Bob performs a quantum measurement
to decode the message.  To ensure security, Alice's operations must meet some conditions.
Reference~\cite{PhysRevA.74.042305} showed that the maximum amount of
information transformed securely equals the quantum mutual information of the
initial state $\rho^{AB}$. This quantum protocol has drawn much attention
and has been further studied~\cite{PhysRevLett.108.040504, PhysRevLett.124.050503}.

Recently, a kind of quantum state named the anyonic state in a two-dimensional
topologically ordered
phase~\cite{PhysRevB.41.9377, PhysRevB.31.2529, witten1989quantum, Bei2019quantum}
has attracted the attention of theorists and experimentalists for its potential in
building a topological fault-tolerant quantum computer~\cite{kitaev2003fault}.
Moreover, the information encoded in anyons presents some unusual properties. In
particular, when a two-dimensional gapped system stays in a topologically ordered
phase, the entanglement entropy of its ground state contains a sub-leading constant
term in addition to the leading area-law
term~\cite{kitaev2006topological, levin2006detecting}. This motivates us to consider
the question of whether the exotic properties of anyonic systems could lead to unexpected
results in the field of quantum communication.

However, the answer to this question is not straightforward. Compared with the Hilbert
space of the conventional quantum system, the Hilbert space of the anyonic system
possesses remarkable properties. The Hilbert space of an anyonic system is decomposed
into a direct sum of sectors
\begin{equation}
  \mathcal{H} = \oplus_q \mathcal{H}_q,
\end{equation}
where $q$ is the charge of a sector. If a local anyonic system is well isolated
from the other anyons, then its charge $q$ will be conserved under any local
physical process. This property is called the superselection rule for an anyonic
system~\cite{PhysRevA.69.052326}, which lays the foundation for its application
in fault-tolerant quantum computing. Moreover, the Hilbert space
of two local anyonic subsystems is not the
direct product of the Hilbert spaces of the two subsystems but a direct sum of some
factorizable Hilbert spaces. To be specific, the Hilbert space of an anyonic system
with the total charge $c$ containing two local subsystems $A$ and $B$ is
\begin{equation}
  \mathcal{H}^c_{AB} = \bigoplus_{a, b} \mathcal{H}_A^a \otimes 
  \mathcal{H}_B^b \otimes V_{ab}^c,
\end{equation}
where $a$ ($b$) is the total charge of subsystem $A$ ($B$) and $V_{ab}^c$ is
the fusion Hilbert space associated with the process that two anyons with
charges $a$ and $b$ fuse into an anyon with charge $c$~\cite{PhysRevA.69.052326}.
Therefore, our existing concepts and methods in quantum information
theory~\cite{nielsen2002quantum} cannot be directly applied
in anyonic systems.

In this paper, we study the deterministic quantum one-time pad, where Bob can
deterministically distinguish different messages, in anyonic systems. For
simplicity we consider the Fibonacci anyon model~\cite{trebst2008short},
the simplest one that can be used for universal topological quantum computation.
We find that the Fibonacci particle-antiparticle pair produced from vacuum,
which was used for anyonic teleportation~\cite{PhysRevLett.101.010501},
\begin{equation}
  \ket{\mathcal{B}} = \ket{\tau, \tau; 1},
\end{equation}
for which the notation will be introduced in Sec.~\ref{sec:one-timepad}, can be used
to transmit messages securely in the asymptotic sense. We will show that the
channel capacity of the state $\ket{\mathcal{B}}$ in the quantum one-time pad equals
the state's anyonic mutual information based on anyonic von Neumann
entropy~\cite{bonderson2017anyonic}, which is a direct generalization of the
corresponding result in Ref.~\cite{PhysRevA.74.042305}. Further more, we
study the deterministic quantum one-time pad via a parameterized state of six
Fibonacci anyons with the total charge being vacuum, which includes the above
three copies of $\ket{\mathcal{B}}$ as a special case. To our surprise, although the three
copies of $\ket{\mathcal{B}}$ have the maximum anyonic mutual information, it is not the
state that sends the maximum number of messages. In the end, we conclude by giving
a summary and an outlook.

\section{Preliminaries}

Before studying the deterministic quantum one-time pad via Fibonacci anyons, we
will introduce the necessary concepts of the quantum one-time pad and the
Fibonacci anyon model.

\subsection{Quantum one-time pad}

In the one-time pad encryption scheme~\cite{Stallings2011,nielsen2002quantum},
Alice and Bob share privately a random key, i.e., a random string of bits. For
every message, Alice encrypts her message by adding the message and the key
using bitwise addition modulo $2$ and sends the encrypted message to Bob publicly.
Then Bob decrypts the encrypted message by subtracting the shared key
to invert the encoding. The message is secure if every random key is
secure and used only once.

The quantum one-time pad proposed by Schumacher and
Westmoreland~\cite{PhysRevA.74.042305} is a quantum analog of the one-time pad,
where a random key shared between Alice and Bob is replaced by a quantum state
$\rho^{AB}$. In this quantum protocol, Alice and Bob initially share a quantum
state $\rho^{AB}$. Alice encrypts her message $\alpha$ by performing a quantum operation
$ {\mathcal{E}}^A_\alpha \otimes I^{B}$ on the state $\rho^{AB},$ and sends her subsystem
$A$ to Bob publicly. Then Bob decrypts the message $\alpha$ by performing a quantum
measurement on the whole system in the state
$\sigma^{AB}_{\alpha}=\mathcal{E}^A_\alpha \otimes I^{B}(\rho^{AB})$. It should be noted that the
subsystem $A$ may also be available to an eavesdropper named Eve. To ensure the
security, Alice's operation must meet the security condition
${\mathcal{E}}^A_\alpha (\rho^A) = \rho^A$ for every $\alpha$, with $\rho^{A}$ being the reduced
states of $\rho^{AB}$.

In general, Alice's task in the quantum one-time pad is to send a classical
message $\alpha$ from some message set with probability $p_\alpha$. Then Bob's 
accessible information is upper bounded by the Holevo
bound~\cite{nielsen2002quantum}
\begin{equation}
  \mathcal{X}(\sigma^{AB}) = S(\sigma^{AB}) - \sum_\alpha p_\alpha S(\sigma_\alpha^{AB}),
\end{equation}
where $S(\eta)$ is the von Neumann entropy of the state $\eta$, i.e.,
$S(\eta)\equiv-\Tr(\eta\log_2 \eta)$. In Ref.~\cite{PhysRevA.74.042305}, this maximum amount of
information transformed securely was proved to equal the quantum mutual
information of the initial state $\rho^{AB}$ asymptotically
\begin{equation}
  \label{eq:13}
  I(\rho^{AB}) = S(\rho^{A}) + S(\rho^{B}) - S(\rho^{AB}).
\end{equation}

For simplicity and convenience, we will restrict ourselves to a specific type of
quantum one-time pad protocol called the deterministic quantum one-time pad
(DQOTP), where the following three assumptions are assumed. First, the key state
$\rho^{AB}$ is a pure state $|\psi\rangle$. Second, all of Alice's encoding operations
$\{\mathcal{E}_{\alpha}^{A}\}$ are unitary transformations $\{U_{\alpha}^{A}\}$. Third,
Bob can distinguish different messages $\{\alpha\}$ by making a quantum measurement,
which implies that the states $\{U_{\alpha}^{A}\otimes I^{B}|\psi\rangle\}$ are orthogonal to each other.

\subsection{Fibonacci anyon model}

In the Fibonacci anyon model, there are two species of anyons: $1$ and $\tau$,
where $1$ denotes the vacuum and $\tau$ denotes the Fibonacci anyon. The species of
anyons are also called topological charges. These anynos can be combined or
split according to the following fusion rules:
\begin{align}\label{eq:fusion}
  1 \times \tau & = \tau, \nonumber\\
  \tau \times 1 & = \tau, \nonumber\\
  \tau \times \tau & = 1 + \tau.
\end{align}
For example, the fusion rule $\tau \times \tau = 1 + \tau$ means there are two possible fusion
results $1$ and $\tau$ when two $\tau$s are fused. It also means charge $1$ ($\tau$)
can be split into two $\tau$s.

The quantum state of the Fibonacci anyon model is defined by the above fusion
rules. For example, the state of $2$ $\tau$s fusing into the vacuum is given by
$\bra{\tau, \tau; 1}$, the state of the vacuum splitting into $2$ $\tau$s is given by
$\ket{\tau, \tau; 1}$. When there are more anyons in the system, we need to specify
the order of fusion or splitting.

In the Fibonacci anyon model, a basic unitary transformation $R_{\tau\tau}$ is
realized by braiding two $\tau$s. By performing a series of such braidings, the
Fibonacci anyon model was shown to be able to realize universal topological
quantum computing~\cite{preskill1999lecture, CMP2002Freedman, PhysRevLett.95.140503}.

For completeness, we give more technical details of the description of the
Hilbert space of the Fibonacci anyon model in the Appendix.

\section{\label{sec:one-timepad}Anyonic deterministic one-time pad}

In this section, we aim to study the deterministic quantum one-time pad via the
Fibonacci anyons. First, we introduce the basic framework for the deterministic
quantum one-time pad. Second, we obtain the average maximal information in the
deterministic quantum one-time pad via Fibonacci Bell states in the asymptotic
limit. Third, we obtain the maximal information in the deterministic quantum
one-time pad via a parameterized state of six Fibonacci anyons.

\subsection{Deterministic quantum one-time pad protocol}\label{subsec:dotp}

Now we are ready to state the protocol of
DQOTP by using the Fibonacci anyonic state $|\psi\rangle$ of $2n$ $\tau$'s with trivial
total charge, where the $n$ $\tau$'s on the left side are owned by Alice, and the
$n$ $\tau$'s on the right side are owned by Bob. Alice performs a unitary
transformation $U^{A}_{\alpha}$ on her $n$ $\tau$'s to encode a message $\alpha$. In
particular, we require that the message $\alpha$ is secure before Bob receives
Alice's $n$ $\tau$'s, which requires that Alice's local state is invariant under the
unitary transformation, i.e.,
\begin{equation}
  \label{eq:secure}
  U^A_{\alpha} \tilde{\rho}^A U^{A \dagger}_{\alpha} = \tilde{\rho}^A,
\end{equation}
where Alice's local state
$\tilde{\rho}^A = \tilde{\rm Tr}[\ket{\psi} \bra{\psi}]$. Then she sends her $n$
$\tau$'s to Bob such that Bob can make a measurement on the $2n$ $\tau$'s to decode the
message $\alpha$. Here we require that Bob can distinguish different messages from
Alice deterministically, i.e., the different unitary transformations satisfying
the orthogonal condition
\begin{equation}
  \label{eq:ortho}
  \bra{\psi} U^{A \dagger}_{\alpha} U^A_{\alpha^{\prime}} \ket{\psi} = 0
\end{equation}
for any different messages $\alpha$ and $\alpha^{\prime}$. Denote the set of unitary
transformations satisfying Eqs.~\eqref{eq:secure} and \eqref{eq:ortho} as
$\mathcal{U}^{A}=\{U^{A}_{\alpha}\}$. Since each
$U^{A}_{\alpha}\in \mathcal{U}^{A}$ represents one distinguishable message for Bob, the
cardinality of $\mathcal{U}^{A}$, denoted as $\abs{\mathcal{U}^{A}}$, equals the
number of such messages sent from Alice to Bob. The central problem in the
DQOTP is to find out the maximum number of  messages from Alice that
Bob can distinguish deterministically. In other words, we shall maximize the
cardinality $\abs{\mathcal{U}^{A}}$, i.e.,
\begin{equation}
  \label{eq:1}
  \max \abs{\mathcal{U}^{A}} \equiv N_{m}(|\psi\rangle).
\end{equation}

\subsection{DQOTP via anyonic Bell states}
\label{sec:dqotp-via-anyonic}

As we know, the Bell state is of great importance in superdense coding,
which is used to send $2$ bits of classical information by sending one qubit.
Then it is natural to ask which state in the Fibonacci anyon system plays a
role similar to the Bell state in the qubit system. A reasonable candidate
for a Bell state in the Fibonacci anyon system~\cite{PhysRevLett.101.010501}
is the state of $2$ $\tau$'s from vacuum
\begin{equation}\label{eq:18}
  \ket{\mathcal{B}} = \ket{\tau, \tau; 1}.
\end{equation}
Then Alice and Bob each have one anyon, and their reduced states are
$\tilde{\rho}_{A}=\tilde{\rho}_{B}=\frac{1}{d_{\tau}} \ket{\tau, 1; \tau} \bra{\tau, 1; \tau}$,
which can be obtained by using the quantum trace and $F$ matrix defined in the Appendix.
The anyonic mutual information for the state $|\mathcal{B}\rangle$ is
$\tilde{\mathcal{I}}(|\mathcal{B}\rangle)=2\log_2 d_{\tau}$, where the anyonic mutual
information for the state $\tilde{\rho}^{AB}$ is defined by
\begin{equation}
  \label{eq:2}
  \tilde{\mathcal{I}}(\rho^{AB}) \equiv \tilde{S}(\tilde{\rho}^A) + \tilde{S}(\tilde{\rho}^B)
    - \tilde{S}(\tilde{\rho}^{AB}),
\end{equation}
where $\tilde{S}(\tilde{\rho}) = - \tilde{\rm Tr} [\tilde{\rho} {\rm log}_2 \tilde{\rho}]$
is the anyonic von-Neumann entropy~\cite{bonderson2017anyonic}
and the quantum trace $\tilde{\rm Tr}$ is defined by Eq.~(\ref{eq:qtrace}). Thus we expect
that the anyonic state $|\mathcal{B}\rangle$ can be used to send $2\log_2 d_{\tau}$ bits of
classical information from Alice and Bob in the anyonic DQOTP\@. However,
Alice has only one $\tau$, and she can not perform any unitary transformation by
braiding, which implies that she can not send classical information in DQOTP\@,
i.e., $N_{m}(|\mathcal{B}\rangle)=0$.

How do we reconcile the above paradox? We turn to study DQOTP by using the
$n$ copies of the state $|\mathcal{B}\rangle$ shared between Alice and Bob
\begin{align}
  \ket{\mathcal{B}}^{\otimes n} & = \ket{\tau, \tau; 1}^{\otimes n} \nonumber\\
  & = \left( \frac{1}{d_\tau} \right)^{\frac{n}{2}}
    \begin{tikzpicture}[baseline]
      \draw (-0.5,1)--(0,0.5)--(0.5,1);
      \draw (-1,1)--(0,0)--(1,1);
      \draw (-2,1)--(0,-1)--(2,1);
      \draw[dotted] (0,-0.3)--(0,-0.7);
      \draw[dotted] (-1.7,1)--(-1.3,1);
      \draw[dotted] (1.3,1)--(1.7,1);
      \draw [decorate,decoration={brace,amplitude=10pt},xshift=0pt,yshift=6pt]
      (-2,1)-- node[midway,yshift=0.6cm]{$n$ $\tau$s}(-0.5,1);
    \end{tikzpicture},
\end{align}
where the first $n$ $\tau$'s from left to right belong to Alice, and the rest belongs
to Bob. Here, we utilize the diagrammatic representation of the anyonic state (see the Appendix).
We aim to show that the state $|\mathcal{B}\rangle$ per copy in the asymptotic
limit can send $2\log_2 d_{\tau}$ bits of classical information in the anyonic DQOTP\@, i.e.,   
\begin{equation}
  \label{eq:3}
  \lim_{n\to \infty}\frac{\log_2 N_{m}(|\mathcal{B}\rangle^{\otimes n})}{n} =
  \tilde{\mathcal{I}}(|\mathcal{B}\rangle)=2\log_2 d_{\tau}. 
\end{equation}

We prove one of our main results, Eq.~\eqref{eq:3}, as follows. By using the
basis transformation named the $F$ move~\cite{bonderson2017anyonic}
\begin{align}
    & \begin{tikzpicture}
        \draw (-1,1)--(0,0) node[pos=0, above]{$a_1$}--(1,1)
          node[pos=1, above]{$\overline{a}_1$};
        \draw (-2,1)--(0,-1) node[pos=0, above]{$a_2$}--(2,1)
          node[pos=1, above]{$\overline{a}_2$};
      \end{tikzpicture} \nonumber \\
    = & \sum_b \left[ F^{a_2 a_1 \overline{a}_1}_{a_2} \right]_{1b}
    \begin{tikzpicture}[baseline]
      \draw (-1,1)--(-0.5,0.5) node[pos=0, above]{$a_1$}--(-1,0);
      \draw (-2,1)--(-1,0) node[pos=0, above]{$a_2$}--(0,-1) node[pos=0.6,above]{$b$}
        --(1,0) node[pos=0.4,above]{$\overline{b}$}--(2,1)
        node[pos=1, above]{$\overline{a}_2$};
      \draw (1,1)--(0.5,0.5) node[pos=0, above]{$\overline{a}_1$}--(1,0);
    \end{tikzpicture},
\end{align}
where the coefficient
$\left[ F^{a_2 a_1 \overline{a}_1}_{a_2} \right]_{1b} =
\sqrt{d_b/(d_{a_1}d_{a_2})}$, we can perform the Schmidt decomposition on the
state $\ket{\mathcal{B}}^{\otimes n}$,
\begin{align}\label{eq:71}
  & \ket{\mathcal{B}}^{\otimes n} \nonumber\\
  = & \left( \frac{1}{d_\tau} \right)^{\frac{n}{2}}
                            \sum_{i=1}^{F(n-1)} \ket{i; n\; \tau \text{s} ; 1}_{A} \otimes \ket{i; n\; \tau \text{s} ; 1}_{B} \nonumber\\
  & + \left( \frac{1}{d_\tau} \right)^{\frac{n-1}{2}} \sum_{j=1}^{F(n)} \ket{j; n\; \tau \text{s} ; \tau}_{A}
                            \otimes \ket{j; n\; \tau \text{s} ; \tau}_{B} \otimes \ket{\tau, \tau; 1},
\end{align}
where $\ket{i; n\; \tau \text{s};a}_{A}$ ($a\in\{1,\tau\}$) is the $i$th orthonormal basis vector of
$n$ $\tau$'s in subsystem $A$ with total charge $a$, the diagrammatic representations for
$\ket{i; n\; \tau \text{s}; a}_{A}$ and $\ket{i; n\; \tau \text{s}; a}_{B}$ are mirror symmetric, and
\begin{equation}\label{eq:Binet}
  F(n) = \frac{d_\tau^n - (-d_\tau)^{-n}}{2d_\tau - 1}
\end{equation}
is Binet's formula~\cite{ball2003strange} which gives the $n$th term in the Fibonacci sequence,
e.g., $F(0) = 0$, $F(1) = 1$, $F(2) = 1$, and $F(3) = 2$. Thus, the local
state of Alice can be written as
\begin{align}
  \label{eq:23}
  \tilde{\rho}_{n}^A = \begin{pmatrix}
      \left( \frac{1}{d_\tau} \right)^n I^1_{F(n-1)} & 0 \\
      0 & \left( \frac{1}{d_\tau} \right)^{n-1} \frac{1}{d_\tau} I^\tau_{F(n)}
    \end{pmatrix},
\end{align}
where $I^a_{d}$ $(a\in\{1,\tau\})$ is the $d\times d$ identity anyonic matrix of space of
$n$ $\tau$'s with total charge $a$. Note that the state $\tilde{\rho}^A_{n}$ can also
be written as $\frac{1}{d_{\tau}^{n}} I_{n \tau}$, where $I_{n \tau}$ is the identity
matrix of the anyonic Hilbert space of $n$ $\tau$'s. 

Now the superselection rule implies that any unitary operator $U^{A}_{\alpha}$
encoding the message $\alpha$ can be decomposed into a direct sum of two sectors,
\begin{align}
  U^A_\alpha = \begin{pmatrix}
      U_\alpha^1 & 0 \\
      0 & U_\alpha^\tau \\
    \end{pmatrix},
\end{align}
where $U_\alpha^1$ ($U_\alpha^\tau$) is the unitary operator living in
sector $1$ (sector $\tau$) of operator $U^A_\alpha$, which meet the security
conditions~\eqref{eq:secure} naturally. Because any two states in
$\{U_{\alpha}^{A}|\mathcal{B}\rangle^{\otimes n}\}$ are orthogonal~\eqref{eq:ortho}, the
cardinality of $\mathcal{U}^{A}$ is not more than the dimension of the Hilbert
space $V_{1}^{2n \tau}$, i.e.,
\begin{equation}
  \label{eq:4}
  N_{m}(|\mathcal{B}\rangle^{\otimes n}) \le F(n-1)^2 + F(n)^2. 
\end{equation}

The orthogonal condition~\eqref{eq:ortho} can be expressed further as $\forall$
different messages $\alpha$ and $\alpha^{\prime}$,
\begin{equation}\label{eq:26}
  \tilde{\rm Tr} \left[ U^{1 \dagger}_{\alpha} U^1_{\alpha^{\prime}} \right] + \tilde{\rm Tr}
    \left[ U^{\tau \dagger}_{\alpha} U^\tau_{\alpha^{\prime}} \right] = 0.
\end{equation}
Then we can construct one $\mathcal{U}^{A}$, denoted by $\mathcal{U}_{L}^{A}$,
that satisfies
\begin{equation}
  \label{eq:5}
    \tilde{\rm Tr} \left[ U^{1 \dagger}_{\alpha} U^1_{\alpha^{\prime}} \right] = \tilde{\rm Tr}
    \left[ U^{\tau \dagger}_{\alpha} U^\tau_{\alpha^{\prime}} \right] = 0.
\end{equation}
Thus, we can find out $\{U^{1}_{\alpha}\}$ and $\{U^{\tau}_{\alpha}\}$ respectively. It is
instructive to consider the DQOTP for the state
\begin{equation}
  \label{eq:6}
  |\phi^{1}\rangle = \frac{1}{\sqrt{F(n-1)}} \sum_{i=1}^{F(n-1)} \ket{i; n\; \tau' \text{s} ;
    1}_{A} \otimes \ket{i; n\; \tau' \text{s} ; 1}_{B}. 
\end{equation}
Obviously, $\{U_{\alpha}^{1}\}$ satisfying Eq.~\eqref{eq:5} is the set of unitary
transformations encoding messages $\mathcal{U}^{A}_{L1}$. In this subspace, the
problem is reduced to the traditional superdense coding for the maximally
entangled state in the $F(n-1)$ dimensional Hilbert space. Thus we can send
$F(n-1)^{2}$ messages from Alice to Bob, which implies that the cardinality
$\abs{\mathcal{U}^{A}_{L1}}=F(n-1)^{2}$. Similarly for
$\mathcal{U}^{A}_{L\tau}=\{U_{\alpha}^{\tau}\}$, the cardinality
$\abs{\mathcal{U}^{A}_{L \tau}}=F(n)^{2}$. Because
$\abs{\mathcal{U}^{A}_{L1}}\le\abs{\mathcal{U}^{A}_{L\tau}}$, we can always take
$\abs{\mathcal{U}^{A}_{L1}}$ elements from $\mathcal{U}^{A}_{L1}$ and $\mathcal{U}^{A}_{L\tau}$
to construct the set $\mathcal{U}^{A}_{L}$, which implies that
$\abs{\mathcal{U}^{A}_{L}}=F(n-1)^{2}$. Thus we obtain
\begin{equation}
  \label{eq:7}
  N_{m}(|\mathcal{B}\rangle^{\otimes n}) \ge \abs{\mathcal{U}^{A}_{L}} = F(n-1)^{2}.
\end{equation}

Then we have the inequality
\begin{equation}
  F(n-1)^2 \le  N_{m}(|\mathcal{B}\rangle^{\otimes n}) \le
  F(n-1)^2 + F(n)^2.
\end{equation}
By using Binet's formula [see Eq.~(\ref{eq:Binet})] we obtain
\begin{equation}\label{eq:30}
  \lim_{n\to \infty} \frac{\log_2 N_{m}(|\mathcal{B}\rangle^{\otimes n})}{n} = 2 {\rm log}_2 d_\tau,
\end{equation}
which completes the proof of Eq.~\eqref{eq:3}.

The main result in Eq.~\eqref{eq:3} reconciles the paradox discussed: Although a
single copy of the anyonic Bell state $|\mathcal{B}\rangle$ can not be used to send
any classical information in the DQOTP\@, one copy in the ensemble of the
anyonic Bell states can send $2\log_2 d_{\tau}$ bits of classical information, which
is equal to the anyonic mutual information of an anyonic Bell state. Thus the
DQOTP protocol gives an operational meaning to anyonic mutual information
and in a sense proves the reliability of the definition of the anyonic entropy.
Since the amount of information relates only to the quantum dimension $d_\tau$ of
Fibonacci anyons, only non-Abelian anyons contain this kind of information.

\subsection{\label{sec:abs} DQOTP via a parameterized state of six Fibonacci
  anyons}

Although the anyonic Bell state $|\mathcal{B}\rangle$ plays an important role in the
anyonic DQOTP\@, it is necessary to study this quantum protocol with a more general
anyonic state. In this section we consider the following parameterized states of
six Fibonacci anyons in the trivial sector
\begin{align}
  \label{eq:8}
  & |\mathcal{G}(p)\rangle \nonumber\\
  = & \sqrt{p}
                     \ket{\tau,\tau,\tau ; 1}_{A} \otimes \ket{\tau,\tau,\tau ; 1}_{B} \nonumber\\
                   & + \sqrt{\frac{1-p}{2}} \sum_{j=1}^{2} \ket{j;\tau,\tau,\tau ; \tau}_{A}
                     \otimes \ket{j; \tau,\tau,\tau ; \tau}_{B} \otimes \ket{\tau, \tau; 1},
\end{align}
where $p \in \left[ 0, 1 \right]$.
Note that when $p=1/d_{\tau}^{3}$ the above state equals three copies of the
anyonic Bell state $|\mathcal{B}\rangle$, i.e.,
$|\mathcal{G}(1/d_{\tau}^{3})\rangle=|\mathcal{B}\rangle^{\otimes 3}$.
And it can be shown that the state $|\mathcal{B}\rangle^{\otimes 3}$ is the
state with maximal mutual information among all states $|\mathcal{G}(p)\rangle$.

\begin{figure}
  \includegraphics[width=1\columnwidth{}, keepaspectratio]{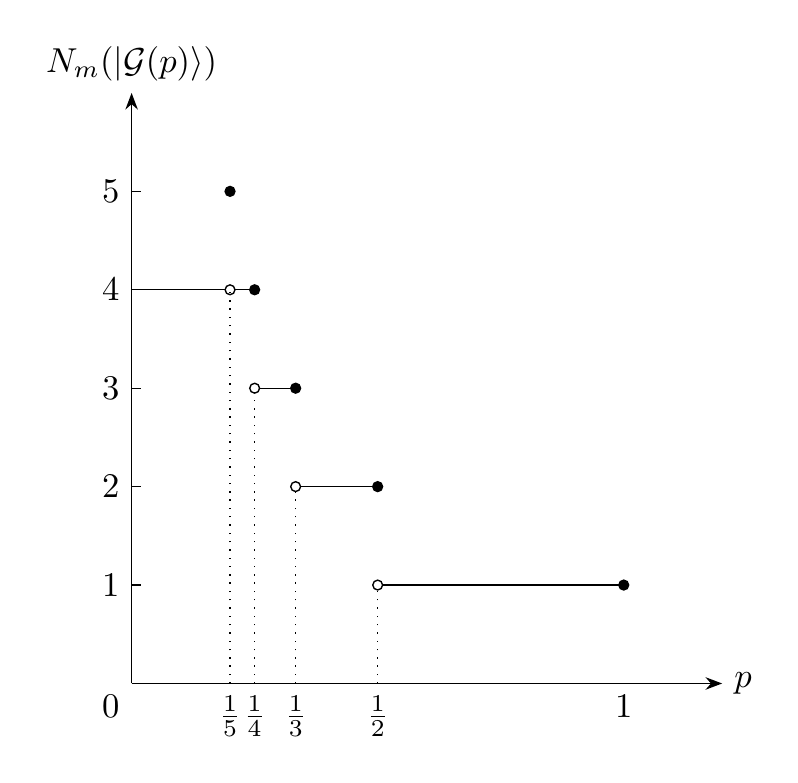}
  \caption{The maximum amount of information $N_{m}(|\mathcal{G}(p)\rangle)$
    transmitted deterministically from Alice to Bob through the state
    $\ket{\mathcal{G}(p)}$ for different $p$.}
  \label{fig:1}
\end{figure}

We aim to show that the maximal amount of information through DQOTP via the
state $|\mathcal{G}(p)\rangle$ is
\begin{equation}
  \label{eq:11}
  N_{m}(|\mathcal{G}(p)\rangle) =
  \begin{cases}
    5 & \text{ if } p = \frac{1}{5}, \\
    4 & \text{ if } 0 \le p < \frac{1}{5} \text{ and } \frac{1}{5} < p \le \frac{1}{4}, \\
    3 & \text{ if } \frac{1}{4} < p \le \frac{1}{3}, \\
    2 & \text{ if } \frac{1}{3} < p \le \frac{1}{2}, \\
    1 & \text{ if } \frac{1}{2} < p \le 1,
\end{cases}
\end{equation}
which is demonstrated in Fig.~\ref{fig:1}. From Eq.~(\ref{eq:11}), we find that
the state $|\mathcal{G}(1/5)\rangle$ that can be used to send the maximal
amount of information is not the maximum entangled state, i.e., three copies of
the anyonic Bell state $|\mathcal{B}\rangle^{\otimes 3}$.

We prove Eq.~\eqref{eq:11} as follows. Note that Alice's local state is
\begin{equation}
  \label{eq:9}
  \tilde{\rho}^{A}(p) =
  \begin{pmatrix}
    p & 0 & 0 \\
    0 & \frac{1-p}{2} \frac{1}{d_\tau} & 0 \\
    0 & 0 & \frac{1-p}{2} \frac{1}{d_\tau}
  \end{pmatrix}.
\end{equation}
Following Eq.~(\ref{eq:secure}), the unitary operator that
Alice uses to encode message $\alpha$ can be written as
\begin{equation}
  U_\alpha^A = \begin{pmatrix}
    1&0&0\\
    0&a_{\alpha}+ib_{\alpha}&c_{\alpha}+id_{\alpha}\\
    0&-c_{\alpha}+id_{\alpha}&a_{\alpha}-id_{\alpha}\\
  \end{pmatrix},
\end{equation}
where the real numbers $a_{\alpha},b_{\alpha},c_{\alpha}$, and $d_{\alpha}$ are functions of
$p$ satisfying $a_{\alpha}^2 + b_{\alpha}^2 + c_{\alpha}^2 + d_{\alpha}^2 = 1$. 
From the viewpoint of geometry, each operator $U_\alpha^A$ can be represented by a
unit vector $\vec{v}_\alpha = (a_{\alpha},b_{\alpha},c_{\alpha},d_{\alpha})$ in four-dimensional Euclidean
space $\mathbb{E}^4$. Now  the orthogonal condition~\eqref{eq:ortho} becomes 
the following inner-product relation in $\mathbb{E}^4$:
\begin{equation}
  \label{eq:10}
  \vec{v}_{\alpha} \cdot \vec{v}_{\alpha^{\prime}} = - \frac{p}{1-p}.
\end{equation}

To obtain the maximal number of messages $N_{m}(|\mathcal{G}(p)\rangle)$ in the
anyonic DQOTP\@, we determine the maximal number of unit vectors
$\{\vec{v}_{\alpha}\}$ by solving Eq.~\eqref{eq:10}, which equals
$N_{m}(|\mathcal{G}(p)\rangle)$. Without loss of generality, we can always take the
first element $U^{A}_{1}$ in the set $\mathcal{U}^A$, which is the $3\times3$ identity
operator, which implies that $\vec{v}_1 = (1, 0, 0 , 0)$. Because the absolute
value of the inner product of any two unit vectors is not more than $1$,
Eq.~\eqref{eq:10} requires that $p\le \frac{1}{2}$. This implies that when
$p>\frac{1}{2}$, no other vector except $\vec{v}_{1}$ exists such
that Eq.~\eqref{eq:10} is satisfied. Thus we find that if $\frac{1}{2}<p\le1$,
then $N_{m}(|\mathcal{G}(p)\rangle)=1$.

Since the inner product of any other unit vector and
the identity $\vec{v}_1$ equals $- \frac{p}{1-p}$, then $\forall \alpha\neq 1$ the unit vector
\begin{equation}
  \vec{v}_{\alpha} = (- \frac{p}{1-p}, b_{\alpha}, c_{\alpha}, d_{\alpha}),
\end{equation}
with $p \in [0, 1/2]$. When $p=\frac{1}{2}$, a unique vector
$\vec{v}_{2}=(-1,0,0,0)$ exists, which implies that
$N_{m}(|\mathcal{G}(\frac{1}{2})\rangle)=2$.

When $0\le p<1/2$, $\forall \alpha\neq 1$, we introduce a unit vector in three-dimensional
Euclidean space $\mathbb{E}_{3}$ by
\begin{equation}
  \vec{w}_{\alpha} = \frac{1-p}{\sqrt{1-2p}} (b_{\alpha}, c_{\alpha}, d_{\alpha}).
\end{equation}
Without loss of generality, we can take $\vec{w}_{2}=(1,0,0)$. For any
$\alpha\ne 1$ and $\alpha^{\prime}\ne 1$, the inner-product relation (\ref{eq:10}) becomes
\begin{equation}\label{eq:angle1}
  \vec{w}_{\alpha} \cdot \vec{w}_{\alpha'} = - \frac{p}{1-2p}.
\end{equation}
Since the absolute value of the above inner product is not more than $1$, we
have $p\le \frac{1}{3}$. In other words, when $\frac{1}{3}<p<\frac{1}{2}$,
only $\vec{w}_{2}$ exists such that Eq.~\eqref{eq:angle1} is satisfied, and
$N_{m}(|\mathcal{G}(p)\rangle)=2$.

When $p=\frac{1}{3}$, Eq.~\eqref{eq:angle1} becomes $\forall \alpha>2$,
$\vec{w}_{2} \cdot \vec{w}_{\alpha}=-1$, i.e., a unique unit vector
$\vec{w}_{3}=(-1,0,0)$ exists. Thus,
$N_{m}(|\mathcal{G}(\frac{1}{3})\rangle)=3$.

When $0\le p<1/3$, $\forall \alpha> 2$, we introduce a unit vector in two-dimensional
Euclidean space $\mathbb{E}_{2}$ by
\begin{equation}
  \vec{x}_{\alpha} = \sqrt{\frac{(1-p)(1-2p)}{1-3p}} (c_{\alpha}, d_{\alpha}).
\end{equation}
Without loss of generality, we can take $\vec{x}_{3}=(1,0)$. For any
$\alpha>2$ and $\alpha^{\prime}>2$, the inner-product relation (\ref{eq:angle1}) becomes
\begin{equation}\label{eq:angle2}
  \vec{x}_{\alpha} \cdot \vec{x}_{\alpha'} = - \frac{p}{1-3p}.
\end{equation}
Since the absolute value of the above inner product is not more than $1$, we
have $p\le \frac{1}{4}$. In other words, when $\frac{1}{4}<p<\frac{1}{3}$,
only $\vec{x}_{3}$ exists such that Eq.~\eqref{eq:angle2} is satisfied, and
$N_{m}(|\mathcal{G}(p)\rangle)=3$.

When $p=\frac{1}{4}$, Eq.~\eqref{eq:angle2} becomes $\forall \alpha>3$, $\vec{x}_{3} \cdot
\vec{x}_{\alpha}=-1$; that is, a unique unit vector
$\vec{x}_{4}=(-1,0)$ exists. Thus, $N_{m}(|\mathcal{G}(\frac{1}{4})\rangle)=4$.

When $0\le p<1/4$, $\forall \alpha> 3$, we introduce a unit vector in two-dimensional
Euclidean space $\mathbb{E}_{1}$ by
\begin{equation}
  y_{\alpha} = \sqrt{\frac{(1-p)(1-3p)}{1-4p}} d_{\alpha}.
\end{equation}
Without loss of generality, we can take $y_{4}=1$. For any
$\alpha>3$ and $\alpha^{\prime}>3$, the inner-product relation (\ref{eq:angle2}) becomes
\begin{equation}\label{eq:angle3}
  y_{\alpha} \cdot y_{\alpha'} = - \frac{p}{1-4p}.
\end{equation}
Since the absolute value of the above inner product is  $1$, we
have $p= \frac{1}{5}$, and $y_{5}=-1$. Thus,
$N_{m}(|\mathcal{G}(\frac{1}{5})\rangle)=5$. In addition, when
$0\le p<\frac{1}{4}$ and $p\neq \frac{1}{5}$, only $y_{4}$ exists such that
Eq.~\eqref{eq:angle3} is satisfied, and $N_{m}(|\mathcal{G}(p)\rangle)=4$. This
completes our proof of Eq.~\eqref{eq:11}.

Note that the above proof is a constructive one. For example, when
$p=\frac{1}{5}$, the five vectors $\{\vec{v}_{\alpha}\}$ obtained are given by
\begin{align*}
  % \label{eq:12}
  \vec{v}_{1} & = (1, 0, 0, 0), \\
  \vec{v}_{2} & = (-\frac{1}{4}, \frac{\sqrt{15}}{4},  0, 0), \\
  \vec{v}_{3} & =  (-\frac{1}{4}, -\frac{\sqrt{15}}{12},  \frac{\sqrt{30}}{6},
                0), \\
  \vec{v}_{4} & = (-\frac{1}{4}, -\frac{\sqrt{15}}{12}, -\frac{\sqrt{30}}{12},
                  \frac{\sqrt{10}}{4}), \\
  \vec{v}_{5} & = (-\frac{1}{4}, -\frac{\sqrt{15}}{12}, -\frac{\sqrt{30}}{12}, -\frac{\sqrt{10}}{4}).
\end{align*}
In geometry, these five vectors construct a regular 5-cell in
four-dimensional Euclidean space $\mathbb{E}^4$. Similarly, for any $p$ the
vectors $\{\vec{v}_{\alpha}\}$ construct
a regular [$N_{m}(|\mathcal{G}(p)\rangle)-1$]-simplex~\cite{coxeter1973regular}.
In other words, the graph corresponds to a regular tetrahedron, a regular triangle,
a segment and a trivial point when $N_{m}(|\mathcal{G}(p)\rangle)$ equals $4$,
$3$, $2$, and $1$, respectively.

How do we understand the results of $N_{m}(|G(p)\rangle)$ from the information
viewpoint? As we know, three copies of the two-qubit Bell state are the maximally
entangled state of the six qubits, which can be used to transmit $6$ bits of
classical information in the superdense coding. The $6$ bits of information are
the maximum amount of the information can be transmitted in the system of six
qubits. Thus, we expect that the same thing happened in the anyonic Bell state
$|\mathcal{B}\rangle$; that is, three copies of the Bell state
$|\mathcal{B}\rangle^{\otimes 3}$ are the state of six $\tau$s that can send the maximum amount
information in the anyonic DQOTP\@. In fact, we find that the state
$|\mathcal{B}\rangle^{\otimes 3}$, i.e., $|\mathcal{G}(1/d_\tau^3)\rangle$,
is the state with maximal mutual information among all the
states $\{|\mathcal{G}(p)\rangle\}$, which seems to support this expectation.
By using Eq.~(\ref{eq:9}), we can obtain the mutual information between $A$
and $B$ of the state $|\mathcal{G}(p)\rangle$,
\begin{align}
  \tilde{I}(|\mathcal{G}(p)\rangle) = & 2 \tilde{S}(\tilde{\rho}^A(p)) \nonumber\\
  = & 2 \left[ -p{\rm log}_2 p - (1-p){\rm log}_2 \frac{1-p}{2 d_\tau} \right].
\end{align}
When $p = 1/d_\tau^3$, $\tilde{I}(|\mathcal{G}(p)\rangle)$ reaches its maximum value.
However, our results in Fig.~\ref{fig:1} show that this is not the case: the
state $|\mathcal{G}(\frac{1}{5})\rangle$ can send five messages, while the state
$|\mathcal{B}\rangle^{\otimes 3}$ can send only four messages.

\begin{figure}
  \includegraphics[width=1\columnwidth{}, keepaspectratio]{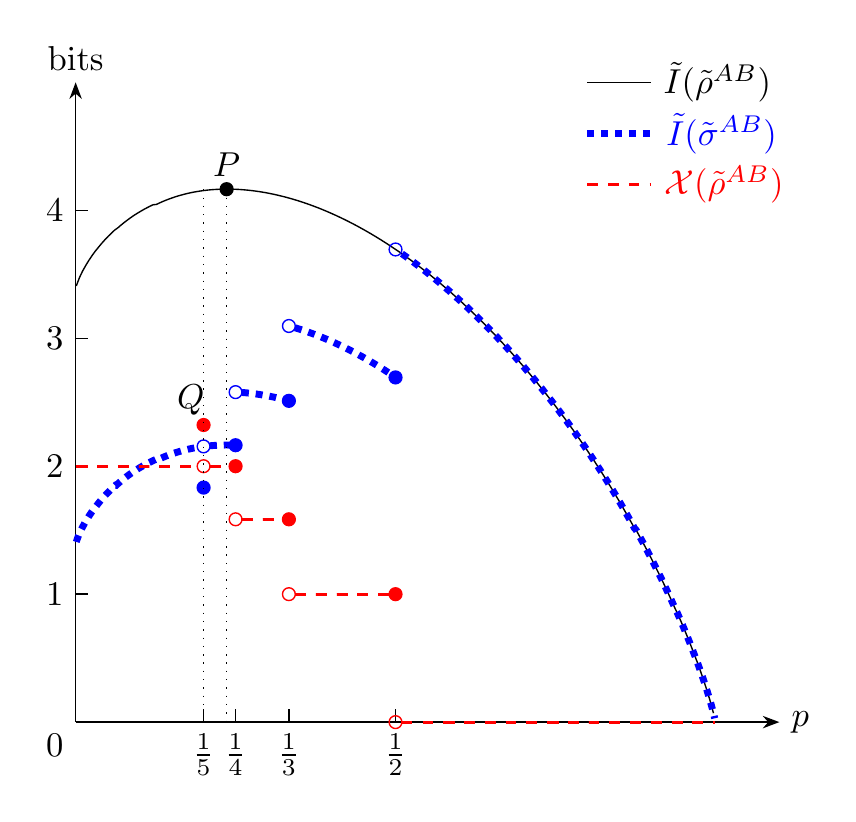}
  \caption{The graph of three functions of $p$: anyonic mutual information
    $\tilde{I}(\tilde{\rho}^{AB})$ (black solid line),
    $\tilde{I}(\tilde{\sigma}^{AB})$ (black dotted line), and the Holevo bound
    $\mathcal{X}(\tilde{\rho}^{AB})$ (red dashed line), which is the difference
    between $\tilde{I}(\tilde{\rho}^{AB})$ and $\tilde{I}(\tilde{\sigma}^{AB})$.
    $\tilde{I}(\tilde{\sigma}^{AB})$ and $\mathcal{X}(\tilde{\rho}^{AB})$ are step
    functions; both are left continuous at points with $p$ coordinates: $1/5$,
    $1/4$, $1/3$, and $1/2$. $P$ ($Q$) is the extreme point of function
    $\tilde{I}(\tilde{\rho}^{AB})$ [$\mathcal{X}(\tilde{\rho}^{AB})$], whose
    $p$ coordinate is $1/d_\tau^3$ ($1/5$). }
  \label{fig:2}
\end{figure}

To resolve the above confusion, it is important to realize that the maximum
amount of information that Alice can transmit to Bob through the state
$\tilde{\rho}^{AB}=\ket{\mathcal{G}(p)}\bra{\mathcal{G}(p)}$ is limited
by the Holevo bound~\cite{nielsen2002quantum},
\begin{align}
  \label{eq:40}
  \mathcal{X}(\tilde{\rho}^{AB}) = & \tilde{S}\left( \sum_\alpha p_\alpha U_\alpha^A \tilde{\rho}^{AB} U_\alpha^{A \dagger} \right) \nonumber\\
                & - \sum_\alpha p_\alpha \tilde{S} \left( U_\alpha^A \tilde{\rho}^{AB}
    U_\alpha^{A \dagger} \right) \nonumber\\
  = & \tilde{S} \left( \tilde{\sigma}^{AB} \right) - \tilde{S} \left(\tilde{\rho}^{AB}\right) \nonumber\\
  = & \left( \tilde{S}(\tilde{\rho}^A) + \tilde{S}(\tilde{\rho}^B) -
      \tilde{S} \left(\tilde{\rho}^{AB}\right) \right) \nonumber\\
  & - \left( \tilde{S}(\tilde{\rho}^A) + \tilde{S}(\tilde{\rho}^B) -
  \tilde{S} \left( \tilde{\sigma}^{AB} \right) \right) \nonumber\\
  = & \tilde{I}(\tilde{\rho}^{AB}) - \tilde{I}(\tilde{\sigma}^{AB}) = \log_2 \abs{\mathcal{U}^{A}},
\end{align}
where
$\tilde{\sigma}^{AB} = \sum_\alpha p_\alpha U_\alpha^A \tilde{\rho}^{AB} U_\alpha^{A \dagger}$, with
$p_{\alpha}=\frac{1}{\abs{\mathcal{U}^{A}}}$. Equation~(\ref{eq:40}) tells us that the
maximum information is limited by the difference between the initial and final
quantum anyonic mutual information, which equals the cardinality of the set of
unitary transformations $\mathcal{U}^{A}$. The results for the
accessible information of $|\mathcal{G}(p)\rangle$, the mutual entropy
$\tilde{I}(\tilde{\rho}^{AB})$, and $\tilde{I}(\tilde{\sigma}^{AB})$ when the cardinality
$\abs{\mathcal{U}^{A}}$ takes its maximum are shown in Fig.~\ref{fig:2},
where the base of the logarithm is $2$.

From Fig.~\ref{fig:2} we can see that although
$\tilde{I}(\ket{\mathcal{B}}^{\otimes 3})$ (point $P$) takes the maximum at
$p=\frac{1}{d_{\tau}^{3}}$ among $\tilde{I}(|\mathcal{G}(p)\rangle)$, the accessible
information $\chi(|\mathcal{G}(p)\rangle)$ takes its maximum at $p=\frac{1}{5}$ (point
$Q$). The reason is that the remaining mutual information
$\tilde{I}(\tilde{\sigma}^{AB})$ at $p=\frac{1}{5}$ is much less than that at
$p=\frac{1}{d_{\tau}^{3}}$. Physically, Alice has five unitary transformations to
send messages by destroying the mutual information at $p=\frac{1}{5}$, while she
has only four such unitary transformations at $p=\frac{1}{d_{\tau}^{3}}$.

\section{Discussion and Conclusion}\label{sec:conclusion}

We investigated the maximum amount of classical information that can be sent
through the DQOTP via Fibonacci anyons. In particular, we obtained two main
results. First, we showed that the anyonic Bell state $\ket{\mathcal{B}}$ can be
used to send $2{\rm log}_2 d_\tau$ bits of classical information asymptotically under
the protocol of DQOTP\@, which is equal to its anyonic mutual information
$\tilde{I}(|\mathcal{B}\rangle)$. Note that a two-qubit Bell state can send $2$ bits
of classical information in the superdense coding. Thus our above result can be
regarded as a generalization of the superdense coding via the two-qubit Bell
state. However, an important difference between them exists: a single copy
of the two-qubit Bell state can send $2$ bits of classical information through
the superdense coding, while a single copy of the anyonic Bell state can not
send classical information through the DQOTP\@, and the anyonic Bell state can
send $2\log_2 d_{\tau}$ bits of classical information per copy only with the aid of
an ensemble of the anyonic Bell states. In other words, Alice must perform some
collective unitary transformations to encode the classical information on all
the anyons on her side, which corresponds to the operations on the Fusion
Hilbert space of her anyons. Furthermore, the above result can be extended from
the Fibonacci anyonic Bell state to the Bell state of any anyon with the
requirement that the anyon model can realize universal topological quantum
computation. An obvious result is that any Bell state of an Abelian anyon can not
send any classical information due to the fact that the dimension of fusion
space of an ensemble of Abelian anyons equals $1$~\cite{PhysRevA.90.062325}.

Second we analytically obtained the maximum number of messages sent in the DQOTP
via a parameterized state $\ket{\mathcal{G}(p)}$ of six Fibonacci anyons, and
gave an informational explanation based on the Holevo bound. We found that
different numbers of messages sent correspond to different regular simplexes in
geometry. In addition we noted that the maximally entangled state
$\ket{\mathcal{B}}^{\otimes 3}$ among all the states $\ket{\mathcal{G}(p)}$ is not the
state that can send the maximum amount of information. In fact, the state
$\ket{\mathcal{B}}^{\otimes 3}$ can be used to send four messages, while the state
$\ket{\mathcal{G}(\frac{1}{5})}$ can send five messages, which is explained by
the accessible information of the state $\ket{\mathcal{G}(\frac{1}{5})}$ being
larger than that of the state $\ket{\mathcal{B}}^{\otimes 3}$.

We presented a framework to study the DQOTP via anyons, and studied two typical
anyonic states to demonstrate the capacity of DQOTP\@. However, there are still
many open problems to investigate. For example, for arbitrary $n$ copies of
$\ket{\mathcal{G}(p)}$ with different values of $p$, what is the maximum amount
information per copy in an ensemble in the DQOTP\@? In addition, how do we deal
with the problems in the DQTOP via an anyon model that can not realize universal
topological quantum computation? Last, but not least, direct experimental
evidence of the Read-Rezayi state arising from the fractional quantum Hall system
at filling fraction $12/5$~\cite{PhysRevB.59.8084}, which is the most promising
candidate for the Fibonacci anyon, has remained elusive. Thus, how to
implement this protocol in the laboratory is a thorny issue.

Our work endows the anyonic von Neumann entropy~\cite{bonderson2017anyonic} with an
operational meaning in the DQOTP, which shows that it may be the right answer to
quantify the information of an anyonic state. Thus we expect that our work might
shed light on all quantum information processes with anyons.

\begin{acknowledgments}
  This work is supported by the NSF of China (Grant No. 11775300 and No. 12075310), the
  National Key Research and Development Program of China (Grant No. 2016YFA0300603), and
  the Strategic Priority Research Program of the Chinese Academy of Sciences
  (Grant No. XDB28000000).
\end{acknowledgments}

\begin{appendix}

\section{Fibonacci anyon model}\label{app:Fanyon}

In this appendix, we briefly review the Fibonacci anyon
model~\cite{kitaev2006anyons, trebst2008short, pachos_2012, bonderson2017anyonic},
focusing on the peculiar structure of its Hilbert space. There are two species
of anyons in the model: $1$ and $\tau$, where $1$ denotes the vacuum and $\tau$
denotes the Fibonacci anyon. The species of anyons are also called topological
charges. These anyons can be combined according to the following fusion rules:
\begin{align}
  1 \times \tau & = \tau, \nonumber\\
  \tau \times 1 & = \tau, \nonumber\\
  \tau \times \tau & = 1 + \tau.
\end{align}
For example, the fusion rule $\tau \times \tau = 1 + \tau$ means there are two possible fusion
results $1$ and $\tau$ when two $\tau$'s are fused. That is why the Fibonacci anyon
$\tau$ is called the non-Abelian anyon.

Based on the above fusion rules, we can define the Hilbert space called the fusion
space, which is spanned by the different fusion paths. For example, the fusion
space of two $\tau$'s fusing into the vacuum is given by
$V_{\tau^2}^1 = {\rm span} \left\{ \bra{\tau, \tau; 1} \right\}$. Similarly the fusion
space of $2$ $\tau$'s fusing into $\tau$ is given by
$V_{\tau^2}^\tau = {\rm span} \left\{ \bra{\tau, \tau; \tau} \right\}$. In particular, each of
these two spaces has only one basis vector since each of them has only one
fusion path. It is useful to employ a diagrammatic representation for anyon
models, where each anyon is associated with an oriented (we will omit the
orientation in this paper) line that can be understood as the anyon's worldline.
In the diagrammatic representation, the two basis vectors above can be
represented as
\begin{align}
  \bra{\tau, \tau; 1} & = \left( \frac{1}{d_\tau} \right)^{\frac{1}{2}}
    \begin{tikzpicture}[baseline]
      \draw[dotted] (0,0)--(0,0.5) node[pos=0.5, right]{$1$};
      \draw (0,0)--(0.5,-0.5) node[pos=1, below]{$\tau$}; 
      \draw (0,0)--(-0.5,-0.5) node[pos=1, below]{$\tau$};
    \end{tikzpicture}, \nonumber\\
  \bra{\tau, \tau; \tau} & = \left( \frac{1}{d_\tau} \right)^{\frac{1}{4}}
    \begin{tikzpicture}[baseline]
      \draw (0,0)--(0,0.5) node[pos=0.5, right]{$\tau$};
      \draw (0,0)--(0.5,-0.5) node[pos=1, below]{$\tau$}; 
      \draw (0,0)--(-0.5,-0.5) node[pos=1, below]{$\tau$};
    \end{tikzpicture},
\end{align}
where $d_\tau = \frac{\sqrt{5} + 1}{2}$ is the quantum dimension of anyon $\tau$,
$\left( 1/d_\tau \right)^{\frac{1}{2}}$ and $\left( 1/d_\tau \right)^{\frac{1}{4}}$
are the normalized coefficients, the solid line denotes $\tau$, and the dotted line
denotes the vacuum (the dotted line will be omitted in the rest of this paper).

For a system with more $\tau$'s, the Hilbert space is constructed by taking the tensor
product of its composite parts. For example, the fusion space $V^\tau_{\tau^3}$
of three $\tau$s with total charge $\tau$ can be constructed as
\begin{equation}
  V^\tau_{\tau^3} \cong \bigoplus_b V^b_{\tau^2} \otimes V^\tau_{b, \tau},
\end{equation}
where $b \in \left\{ 1, \tau \right\}$. It should be noted that fusion order is not
unique. In the example above you can choose to start with fusion of the two $\tau$'s
on the left or the two $\tau$s on the right. These two different methods of fusion
are related by $F$ matrix:
\begin{align}
  \begin{tikzpicture}[baseline=(current bounding box.center)]
    \draw (-1,-1)--(0,0) node[pos=0, below]{$\tau$} node[pos=0.6, above]{$b$}--(1,-1)
      node[pos=1, below]{$\tau$};
    \draw (0,-1)--(-0.5,-0.5) node[pos=0, below]{$\tau$};
    \draw (0,0)--(0,0.5) node[pos=0.5, right]{$\tau$};
  \end{tikzpicture} = \sum_d \left( F^\tau_{\tau \tau \tau} \right)^b_d
    \begin{tikzpicture}[baseline=(current bounding box.center)]
      \draw (-1,-1)--(0,0) node[pos=0, below]{$\tau$} --(1,-1) node[pos=0.4, above]{$d$}
        node[pos=1, below]{$\tau$};
      \draw (0,-1)--(0.5,-0.5) node[pos=0, below]{$\tau$};
      \draw (0,0)--(0,0.5) node[pos=0.5, right]{$\tau$};
    \end{tikzpicture},
\end{align}
where $b,d \in \{ 1, \tau\}$ and 
\begin{align}
  F^\tau_{\tau \tau \tau} =
    \begin{pmatrix}
      \frac{1}{d_\tau} & \frac{1}{\sqrt{d_\tau}} \\
      \frac{1}{\sqrt{d_\tau}} & -\frac{1}{d_\tau} \\
    \end{pmatrix}.
\end{align}
Unless a specific statement about the fusion order is given, we generally adopt
fusion order from left to right. The dimension $D_n$ of fusion space
$V^1_{\tau^n}$ of $n$ $\tau$'s with trivial total charge gives the Fibonacci sequence,
$0, 1, 1, 2, 3, 5, 8, \cdots$, and when $n$ is large, this dimension is approximately
proportional to $d_{\tau}^{n}$, i.e.,
\begin{equation}
  D_n \propto d_\tau^n.
\end{equation}
Thus, one may think of this quantum dimension $d_\tau$ denotes the dimension of the internal
space of Fibonacci anyon $\tau$. The splitting space which is the dual space of fusion
space can be defined in the same way. For example, the splitting space of one $\tau$
splitting into two $\tau$'s is given by
$V^{\tau^2}_1 = {\rm span} \left\{ \ket{\tau, \tau; \tau} \right\}$ and
\begin{align}
  \ket{\tau, \tau; \tau} = \left( \frac{1}{d_\tau} \right)^{\frac{1}{4}}
    \begin{tikzpicture}[baseline]
      \draw (0,0)--(0,-0.5) node[pos=0.5, right]{$\tau$};
      \draw (0,0)--(0.5,0.5) node[pos=1, above]{$\tau$}; 
      \draw (0,0)--(-0.5,0.5) node[pos=1, above]{$\tau$};
  \end{tikzpicture}.
\end{align}

A linear anyonic operator can be defined using the basis vectors in fusion space and
splitting space as we do in quantum mechanics. For example, the identity operator
for two Fibonacci anyons is
\begin{align}
  \mathbbm{1}_{\tau \tau} = & \ket{\tau, \tau; 1} \bra{\tau, \tau; 1}
    + \ket{\tau, \tau; \tau} \bra{\tau, \tau; \tau} \nonumber\\
  = & \frac{1}{d_\tau} \begin{tikzpicture}[baseline]
    \draw (-0.5,0.75)--(0,0.25) node[pos=0,above]{$\tau$}--(0.5,0.75)
      node[pos=1, above]{$\tau$};
    \draw (-0.5,-0.75)--(0,-0.25) node[pos=0, below]{$\tau$}--(0.5,-0.75)
      node[pos=1, below]{$\tau$};
  \end{tikzpicture} + \frac{1}{\sqrt{d_\tau}} \begin{tikzpicture}[baseline]
    \draw (-0.5,0.75)--(0,0.25) node[pos=0,above]{$\tau$}--(0.5,0.75)
      node[pos=1, above]{$\tau$};
    \draw (-0.5,-0.75)--(0,-0.25) node[pos=0, below]{$\tau$}--(0.5,-0.75)
      node[pos=1, below]{$\tau$};
    \draw (0,0.25)--(0,-0.25) node[pos=0.5, left]{$\tau$};
  \end{tikzpicture}.
\end{align}
Since the basis vector of space $V_{\tau^n}^1$ is orthogonal to the basis vector
of $V_{\tau^n}^\tau$, the operator $M$ in the Fibonacci anyon model is decomposed into a
sum of sector $1$ and sector $\tau$: $M = M^1 \oplus M^\tau$, where $M^1$ ($M^\tau$)
is the anyonic operator in sector $1$ (sector $\tau$).

The quantum trace~\cite{kitaev2006anyons,bonderson2017anyonic},
which joins the outgoing anyon lines of the anyonic operator back onto the
corresponding incoming lines, e.g.,
\begin{equation}
  \tilde{\rm Tr} \left[ \frac{1}{\sqrt{d_\tau}} \begin{tikzpicture}[baseline]
    \draw (-0.5,0.75)--(0,0.25) node[pos=0,above]{$\tau$}--(0.5,0.75)
      node[pos=1, above]{$\tau$};
    \draw (-0.5,-0.75)--(0,-0.25) node[pos=0, below]{$\tau$}--(0.5,-0.75)
      node[pos=1, below]{$\tau$};
    \draw (0,0.25)--(0,-0.25) node[pos=0.5, left]{$\tau$};
  \end{tikzpicture} \right] = \frac{1}{\sqrt{d_\tau}}
  \begin{tikzpicture}[baseline]
    \draw (-0.5,0.75)--(0,0.25) node[pos=0,above]{$\tau$}--(0.5,0.75)
      node[pos=1, above]{$\tau$};
    \draw (-0.5,-0.75)--(0,-0.25) node[pos=0, below]{$\tau$}--(0.5,-0.75)
      node[pos=1, below]{$\tau$};
    \draw (0,0.25)--(0,-0.25) node[pos=0.5, left]{$\tau$};
    \draw (0.5,0.75)--(1,0.25)--(1,-0.25)--(0.5,-0.75);
    \draw (-0.5,0.75)--(0.3,1.55)--(1.2,0.65)--(1.2,-0.65)--(0.3,-1.55)--(-0.5,-0.75);
  \end{tikzpicture},
\end{equation}
is defined to be related to the conventional trace by
\begin{equation}\label{eq:qtrace}
  \tilde{\rm Tr} \left[ M \right] =  {\rm Tr} \left[ M^1 \right]
    + d_\tau {\rm Tr} \left[ M^\tau \right],
\end{equation}
where the conventional trace ${\rm Tr}$ of an operator is the sum of diagonal elements,
e.g.,
\begin{equation}
  {\rm Tr} \left[ \ket{a, b; c} \bra{a', b'; c} \right]
    = \delta_{aa'} \delta_{bb'}.
\end{equation}
By using the quantum trace, we can define an operator $\tilde{\rho}$, called an anyonic
density operator satisfying the normalization condition
$\tilde{\rm Tr} \left[ \tilde{\rho} \right] = 1$ and the positive semi-definite
condition; that is, for any anyonic state $\ket{\phi}$, we have
$\tilde{\rm Tr} \left[ \bra{\phi} \tilde{\rho} \ket{\phi} \right] \ge 0$.

In addition to fusion rules, the Fibonacci anyon model also needs to meet the rules of
braiding. Specifically, exchanging neighboring $\tau$s gives to the anyonic state
a unitary evolution named the $R$ matrix:
\begin{align}
  \begin{tikzpicture}[baseline]
    \draw (0,-0.5)--(0,0) node[pos=0.5,right]{$b$};
    \draw (0,0) to [out=160,in=225] (-0.2,0.5)--(0.3,1) node[pos=1,above]{$\tau$};
    \draw (0,0) to [out=20,in=-45] (0.2,0.5)--(0.1,0.6);
    \draw (-0.1,0.8)--(-0.3,1) node[pos=1,above]{$\tau$};
  \end{tikzpicture} & = \sum_{d} \left( R_{\tau \tau} \right)_{b}^{d}
    \begin{tikzpicture}[baseline]
      \draw (0,-0.5)--(0,0) node[pos=0.5,right]{$d$};
      \draw (0,0)--(-0.5,0.5) node[pos=1,above]{$\tau$};
      \draw (0,0)--(0.5,0.5) node[pos=1,above]{$\tau$};
    \end{tikzpicture},
\end{align}
where  $b,d\in\{1, \tau\}$ and
\begin{align}
  R_{\tau \tau} =
        \begin{pmatrix}
          e^{i4\pi/5} & 0 \\
          0 & -e^{i2\pi/5} \\
        \end{pmatrix}.
\end{align}
Using $F$ and $R$ matrices, we can represent any braiding operators. It has been
proved that by using the braiding operators $b_1 = R_{\tau \tau}$ and
$b_2 = (F^\tau_{\tau \tau \tau})^{-1} R_{\tau \tau} F^\tau_{\tau \tau \tau}$, which
denote exchanging the first $\tau$ with the second $\tau$ clockwise and exchanging
the second $\tau$ with the third $\tau$ clockwise respectively, we can simulate any
unitary operators acting on the space formed by three $\tau$'s with arbitrary accuracy.
That's the key reason why the Fibonacci anyon model can be shown to realize
universal topological quantum computation~\cite{preskill1999lecture, CMP2002Freedman}.

\end{appendix}

\bibliographystyle{unsrt}
\bibliography{AHBbib.bib}

\begin{thebibliography}{10}

\bibitem{PhysRevLett.69.2881}
Charles~H. Bennett and Stephen~J. Wiesner.
\newblock Communication via one- and two-particle operators on
  einstein-podolsky-rosen states.
\newblock {\em Phys. Rev. Lett.}, 69:2881--2884, Nov 1992.

\bibitem{PhysRevLett.76.4656}
Klaus Mattle, Harald Weinfurter, Paul~G. Kwiat, and Anton Zeilinger.
\newblock Dense coding in experimental quantum communication.
\newblock {\em Phys. Rev. Lett.}, 76:4656--4659, Jun 1996.

\bibitem{PhysRevA.74.042305}
Benjamin Schumacher and Michael~D. Westmoreland.
\newblock Quantum mutual information and the one-time pad.
\newblock {\em Phys. Rev. A}, 74:042305, Oct 2006.

\bibitem{PhysRevLett.108.040504}
Fernando G. S.~L. Brand\~ao and Jonathan Oppenheim.
\newblock Quantum one-time pad in the presence of an eavesdropper.
\newblock {\em Phys. Rev. Lett.}, 108:040504, Jan 2012.

\bibitem{PhysRevLett.124.050503}
Kunal Sharma, Eyuri Wakakuwa, and Mark~M. Wilde.
\newblock Conditional quantum one-time pad.
\newblock {\em Phys. Rev. Lett.}, 124:050503, Feb 2020.

\bibitem{PhysRevB.41.9377}
X.~G. Wen and Q.~Niu.
\newblock Ground-state degeneracy of the fractional quantum hall states in the
  presence of a random potential and on high-genus riemann surfaces.
\newblock {\em Phys. Rev. B}, 41:9377--9396, May 1990.

\bibitem{PhysRevB.31.2529}
F.~D.~M. Haldane and E.~H. Rezayi.
\newblock Periodic laughlin-jastrow wave functions for the fractional quantized
  hall effect.
\newblock {\em Phys. Rev. B}, 31:2529--2531, Feb 1985.

\bibitem{witten1989quantum}
Edward Witten.
\newblock Quantum field theory and the jones polynomial.
\newblock {\em Communications in Mathematical Physics}, 121(3):351--399, 1989.

\bibitem{Bei2019quantum}
Bei Zeng, Xie Chen, Duan-Lu Zhou, and Xiao-Gang Wen.
\newblock {\em Quantum Information Meets Quantum Matter}.
\newblock Springer-Verlag New York, 2019.

\bibitem{kitaev2003fault}
A~Yu Kitaev.
\newblock Fault-tolerant quantum computation by anyons.
\newblock {\em Ann. Phys. (N.Y.)}, 303(1):2--30, 2003.

\bibitem{kitaev2006topological}
Alexei Kitaev and John Preskill.
\newblock Topological entanglement entropy.
\newblock {\em Phys. Rev. Lett.}, 96:110404, Mar 2006.

\bibitem{levin2006detecting}
Michael Levin and Xiao-Gang Wen.
\newblock Detecting topological order in a ground state wave function.
\newblock {\em Phys. Rev. Lett.}, 96:110405, Mar 2006.

\bibitem{PhysRevA.69.052326}
Alexei Kitaev, Dominic Mayers, and John Preskill.
\newblock Superselection rules and quantum protocols.
\newblock {\em Phys. Rev. A}, 69:052326, May 2004.

\bibitem{nielsen2002quantum}
Michael~A. Nielsen and Isaac~L. Chuang.
\newblock {\em Quantum Computation and Quantum Information: 10th Anniversary
  Edition}.
\newblock Cambridge University Press, Cambridge, 2010.

\bibitem{trebst2008short}
Simon Trebst, Matthias Troyer, Zhenghan Wang, and Andreas~WW Ludwig.
\newblock A short introduction to fibonacci anyon models.
\newblock {\em Progress of Theoretical Physics Supplement}, 176:384--407, 2008.

\bibitem{PhysRevLett.101.010501}
Parsa Bonderson, Michael Freedman, and Chetan Nayak.
\newblock Measurement-only topological quantum computation.
\newblock {\em Phys. Rev. Lett.}, 101:010501, Jun 2008.

\bibitem{bonderson2017anyonic}
Parsa Bonderson, Christina Knapp, and Kaushal Patel.
\newblock Anyonic entanglement and topological entanglement entropy.
\newblock {\em Ann. Phys. (N.Y.)}, 385:399--468, 2017.

\bibitem{Stallings2011}
William Stallings.
\newblock {\em Cryptography and Security, 5th ed.}
\newblock Cambridge University Press, Cambridge, 2011.

\bibitem{preskill1999lecture}
John Preskill.
\newblock Lecture notes for physics 219: Quantum computation.
\newblock {\em Caltech Lecture Notes}, 1999.

\bibitem{CMP2002Freedman}
Michael~H. Freedman, Michael Larsen, and Zhenghan Wang.
\newblock A modular functor which is universal for quantum computation.
\newblock {\em Communications in Mathematical Physics}, 227:605--622, Jun 2002.

\bibitem{PhysRevLett.95.140503}
N.~E. Bonesteel, L.~Hormozi, G.~Zikos, and S.~H. Simon.
\newblock Braid topologies for quantum computation.
\newblock {\em Phys. Rev. Lett.}, 95:140503, Sep 2005.

\bibitem{ball2003strange}
Keith~M Ball.
\newblock {\em Strange curves, counting rabbits, and other mathematical
  explorations}.
\newblock Princeton University Press, Princeton, NJ, 2003.

\bibitem{coxeter1973regular}
H.~S.~M. Coxeter.
\newblock {\em Regular Polytopes: 3d Ed}.
\newblock Dover, New York, 1973.

\bibitem{PhysRevA.90.062325}
Kohtaro Kato, Fabian Furrer, and Mio Murao.
\newblock Information-theoretical formulation of anyonic entanglement.
\newblock {\em Phys. Rev. A}, 90:062325, Dec 2014.

\bibitem{PhysRevB.59.8084}
N.~Read and E.~Rezayi.
\newblock Beyond paired quantum hall states: Parafermions and incompressible
  states in the first excited landau level.
\newblock {\em Phys. Rev. B}, 59:8084--8092, Mar 1999.

\bibitem{kitaev2006anyons}
Alexei Kitaev.
\newblock Anyons in an exactly solved model and beyond.
\newblock {\em Ann. Phys. (N.Y.)}, 321(1):2--111, 2006.

\bibitem{pachos_2012}
Jiannis~K. Pachos.
\newblock {\em Introduction to Topological Quantum Computation}.
\newblock Cambridge University Press, Cambridge, 2012.

\end{thebibliography}

\end{document}